\documentclass{JINST}
\let\ifpdf\relax

\usepackage{lineno}
\usepackage{subfigure}
\usepackage{comment}

\newcommand\lum{\ensuremath{\mathrm{cm}^{-2} \mathrm{s^{-1}}}} 
\newcommand{\neqcm}{\ensuremath{\mathrm{n}_{\mathrm{eq}}/\mathrm{cm}^2}}
\newcommand{\mum}{$\mathrm{\mu}$m}

\newcommand{\degC}{\ensuremath{^{\circ}\mathrm{C}}}

\newcommand{\Sr}{\ensuremath{^{90}}Sr}
\newcommand{\Cd}{\ensuremath{^{109}}Cd}
\newcommand{\Am}{\ensuremath{^{241}}Am}

\title{Thin n-in-p planar pixel sensors and active edge sensors for the ATLAS upgrade at HL-LHC}

\author{S.~Terzo$^a$\thanks{Corresponding author.}~, A.~Macchiolo$^a$, R.~Nisius$^a$ and B.~Paschen$^a$\\
\llap{$^a$}Max-Planck-Institut f\"ur Physik (Werner-Heisenberg-Institut),\\
  F\"ohringer Ring 6, D-80805 M\"unchen, Germany\\
E-mail: \email{Stefano.Terzo@mpp.mpg.de}}

\abstract{Silicon pixel modules employing n-in-p planar sensors with an active thickness of 200~\mum{}, produced at CiS, and 100-200~\mum{} thin active/slim edge sensor devices, produced at VTT in Finland have been interconnected to ATLAS FE-I3 and FE-I4 read-out chips. The thin sensors are designed for high energy physics collider experiments to ensure radiation hardness at high fluences. Moreover, the active edge technology of the VTT production maximizes the sensitive region of the assembly, allowing for a reduced overlap of the modules in the pixel layer close to the beam pipe. The CiS production includes also four chip sensors according to the module geometry planned for the outer layers of the upgraded ATLAS pixel detector to be operated at the HL-LHC. The modules have been characterized using radioactive sources in the laboratory and with high precision measurements at beam tests to investigate the hit efficiency and charge collection properties at different bias voltages and particle incidence angles. The performance of the different sensor thicknesses and edge designs are compared before and after irradiation up to a fluence of 1.4$\times$10$^{16}$~\neqcm{}.}

\begin{document}

\section{Introduction}\label{sec:intro}
The upgrade of the LHC planned to be operative from 2025 onwards (HL-LHC) aims at increasing the instantaneous luminosity of the accelerator up to 5$\times$10$^{34}$~\lum{}~\cite{hl-lhc}. To face the higher particle fluence expected, and to profit from the huge amount of data that the HL-LHC will deliver, an upgrade of the ATLAS detector will be necessary, including a full replacement of the inner pixel detector with new radiation hard devices~\cite{phase2}. The foreseen layout employs 4 barrel layers with a radial distance of the innermost layer from the proton beam line of only 3.9 cm. New planar pixel modules based on the n-in-p pixel technology and employing thin sensors to enhance radiation hardness, have been designed to fulfill the requirements for the HL-LHC upgrade of the ATLAS inner detector. The n-in-p pixel technology has already been proven to be a potentially cost effective alternative to n-in-n devices presently used in ATLAS~\cite{n-in-p,n-in-p-kek}. The prototypes characterized in this paper include four chip sensors to cover large areas in the outer layers as well as active/slim edge sensors designed for the innermost layers, where an overlap of pixel modules along the beam direction is not allowed by space restrictions. The various sensor thicknesses and designs for active edge devices are investigated and compared after high radiation fluences.

\section{Module characterization}\label{sec:concept}
\paragraph{Experimental setups.} The charge collection studies presented in this paper have been performed with \Sr{} $\beta$-electrons using the USBPix readout system~\cite{usbpix}. In addition, measurements performed with \Am{} and \Cd{} $\gamma$ sources have been used as a reference to improve the charge calibration. The modules are tested inside a climate chamber at a stable environmental temperature of $20$~\degC{} before irradiation and at -50~\degC{} after irradiation (corresponding to almost -40~\degC{} on the sensor~\cite{weigell}). The measurements of the hit efficiency have been performed in beam test experiments in the framework of the ATLAS Planar Pixel Sensors (PPS) collaboration both at DESY in Hamburg using 4~GeV electrons and at the SpS at CERN with 120~GeV pions.
The EUDET telescope is used for particle tracking, which allows for a pointing resolution of the telescope on the Device Under Test (DUT) as low as 2~\mum{} in the case of high energetic pions~\cite{EUDET,eff-err}. In case of irradiated devices the DUTs are cooled with dry-ice to a measured temperature on the sensor between -50~\degC{} and -40~\degC{}.
The hit efficiency of a given sensor is defined as the ratio of the number of reconstructed tracks associated to a cluster in the DUT to the total number of tracks reconstructed by the telescope crossing the active area of the DUT. The impact point of a track is calculated in the middle of the thickness of the active bulk, and a cluster is associated to a track if the track crossed at least one of its pixels. An absolute systematic uncertainty of 0.3\% is associated to all hit efficiency measurements in this paper according to~\cite{eff-err}. 

\subsection{The thin active edge pixels production at VTT}\label{sec:active}
\begin{figure*}[tbp] 
\centering
\subfigure[]{
	\includegraphics[width=.46\textwidth]{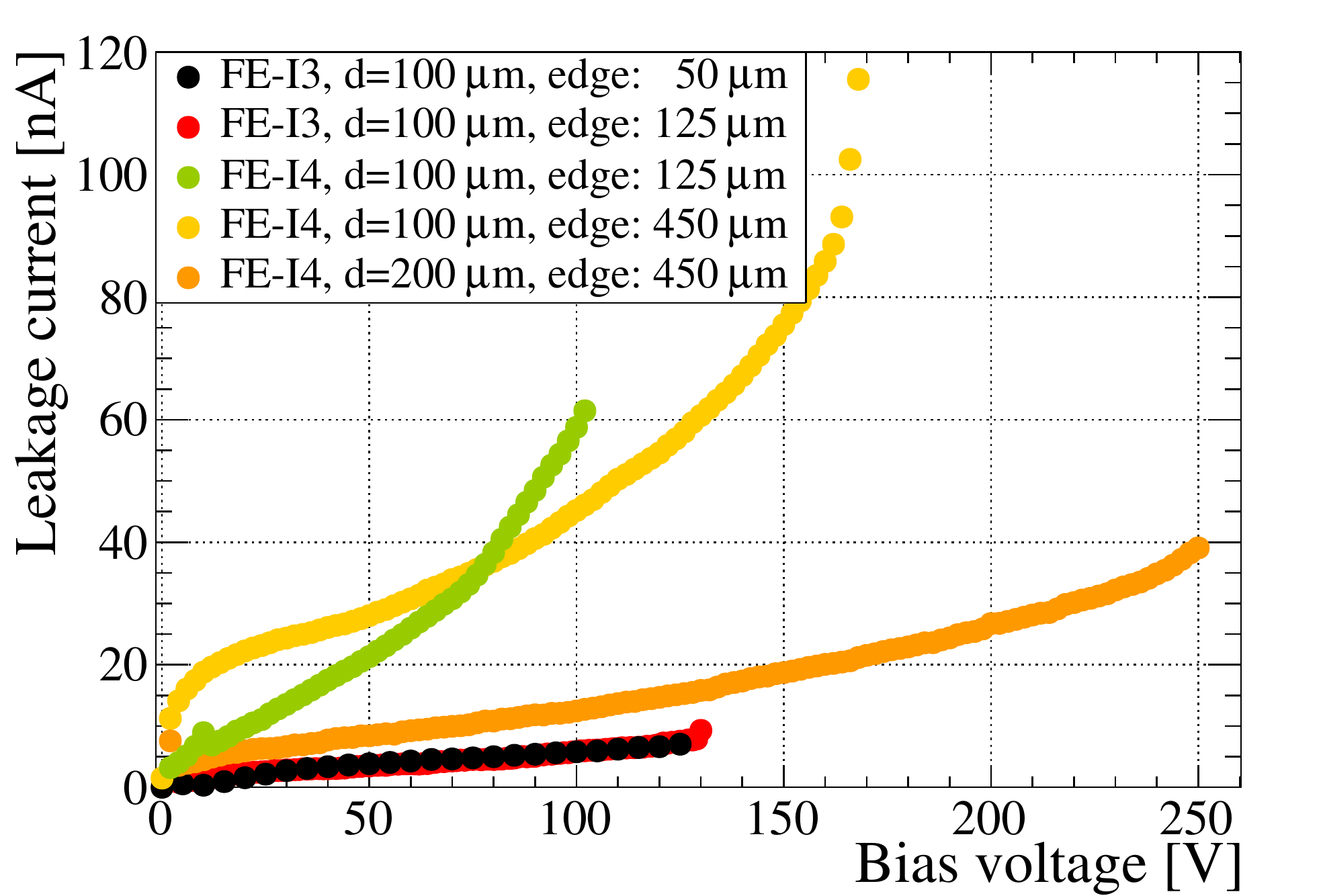}
	\label{fig:VTT-iv_a}
}
\subfigure[]{
	\includegraphics[width=.457\textwidth]{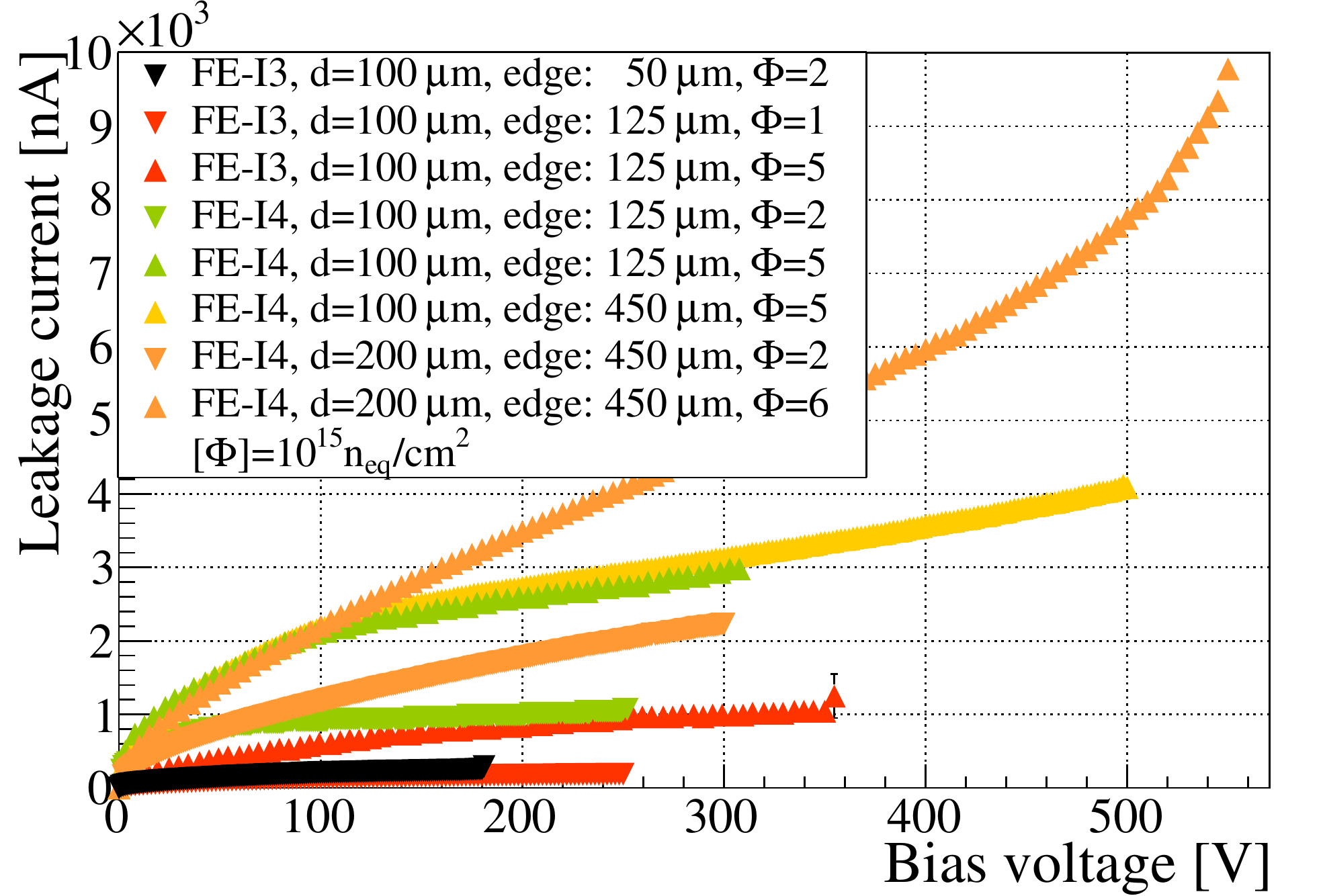}
	\label{fig:VTT-iv_b}
}
\caption{The leakage current as a function of the bias voltage is shown \protect\subref{fig:VTT-iv_a} before and \protect\subref{fig:VTT-iv_b} after irradiation for the different VTT sensor designs.}
\label{fig:VTT-iv}
\end{figure*}

Planar n-in-p pixel sensors with active edges have been produced at VTT Finland, on p-type FZ silicon with an initial resistivity of 10~k$\Omega\,$cm. The sensors are thinned to 100 or 200~\mum{}. The fabrication requires a support wafer that allows for etching trenches at the sensor borders using Deep Reactive Ion Etching (DRIE). The boron implant of the backside of the p-type sensors is then extended to the sides with a four-quadrant ion implantation~\cite{fqii1,fqii2}. Afterwards, the handle wafer is removed and the sensors are interconnected with solder bump bonding to either FE-I3~\cite{fei3} or FE-I4~\cite{fei4} ATLAS chips. Homogeneous p-spray has been used for the inter-pixel isolation. Three different edge designs have been implemented~\cite{slid-tsv-japan}: a classic design with an inactive region between the last pixel implant and the sensor edge of d$_{\mathrm{edge}}=450$~\mum{}, which includes 10 Guard Ring (GR) structures and one Bias Ring (BR); a slim edge design employing only a BR with d$_{\mathrm{edge}}=125$~\mum{}; and finally an active edge design characterized by a distance between the last pixel implant and the activated edge of only d$_{\mathrm{edge}}=50$~\mum{} with just one floating GR. 
\paragraph{Irradiations.} One FE-I3 module, 100~\mum{} thin with d$_{\mathrm{edge}}=50$~\mum{} has been irradiated with neutrons from the nuclear reactor in Ljubljana to a fluence of 2$\times$10$^{15}$~\neqcm{}. Two modules, a FE-I3 assembly with a 100~\mum{} thin slim edge sensor and a 200~\mum{} thick FE-I4 assembly, where exposed to mixed irradiation with 23~MeV protons at KIT and reactor neutrons in Ljubljana up to a total fluence of 5 and 6$\times$10$^{15}$~\neqcm{}, respectively. Results obtained after irradiations to the intermediate fluence are discussed in~\cite{slid-tsv-japan}.
The IV curves before and after irradiation are shown in figure~\ref{fig:VTT-iv}. For the not irradiated devices the breakdown voltages are between 100~V and 140~V and the full depletion voltage is around 15~V, as expected from the high bulk resistivity. After irradiation the sensors with d$_{\mathrm{edge}}=125$~\mum{} show a breakdown voltage between 250 and 350~V that is higher than than the voltage at which the charge saturation for 100~\mum{} thin sensors occurs (figure~\ref{fig:thick-comp}). The sensors with d$_{\mathrm{edge}}=450$~\mum{} and full GR structure are operative up to 500~V. The more aggressive 50~\mum{} active edge design shows a lower breakdown voltage at 180~V at the beginning of the charge saturation.

\begin{figure*}[tbp] 
\centering
\subfigure[]{
	\includegraphics[width=.43\textwidth]{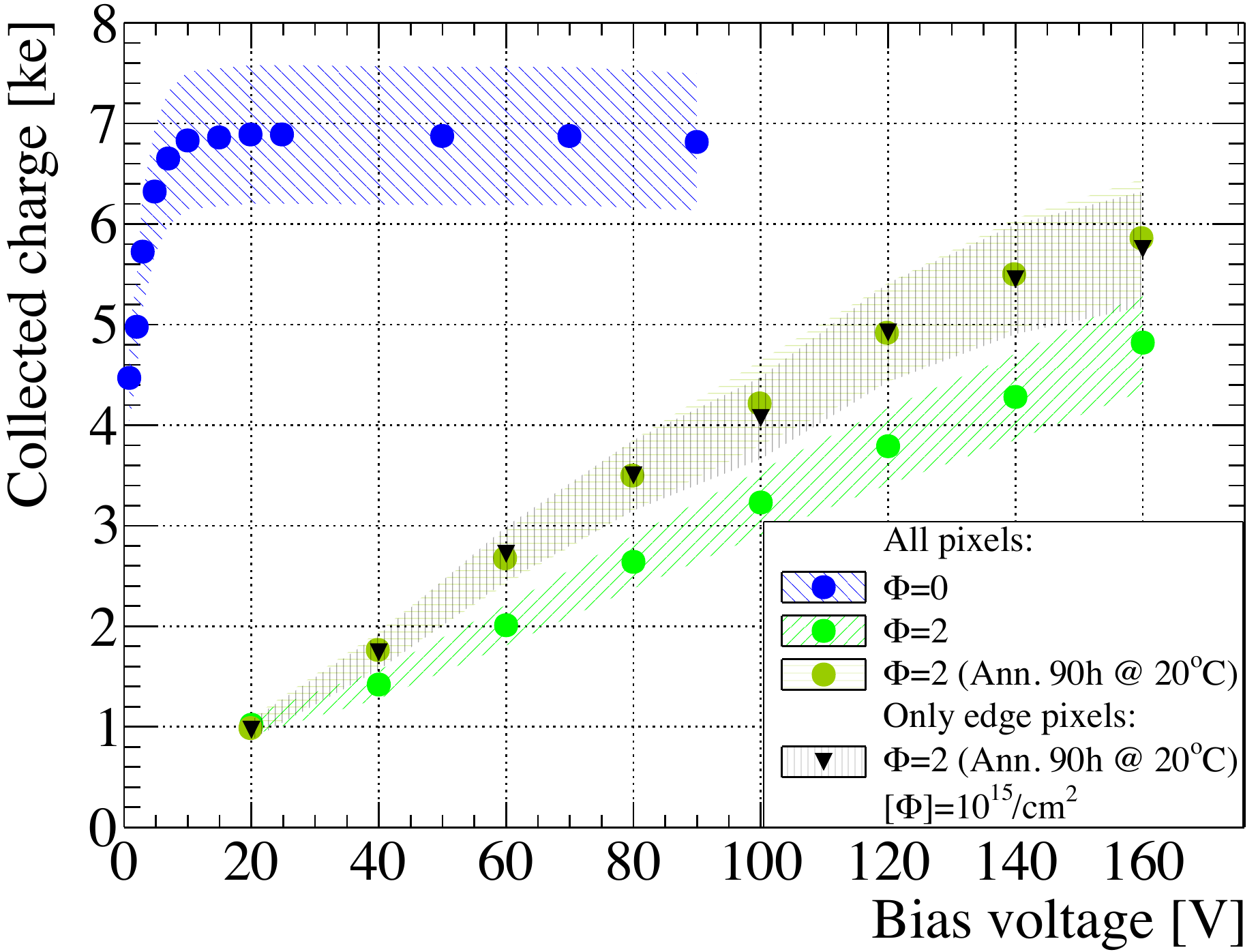}
	\label{fig:VTT-edge-50-a}
}
\subfigure[]{
	\includegraphics[width=.47\textwidth]{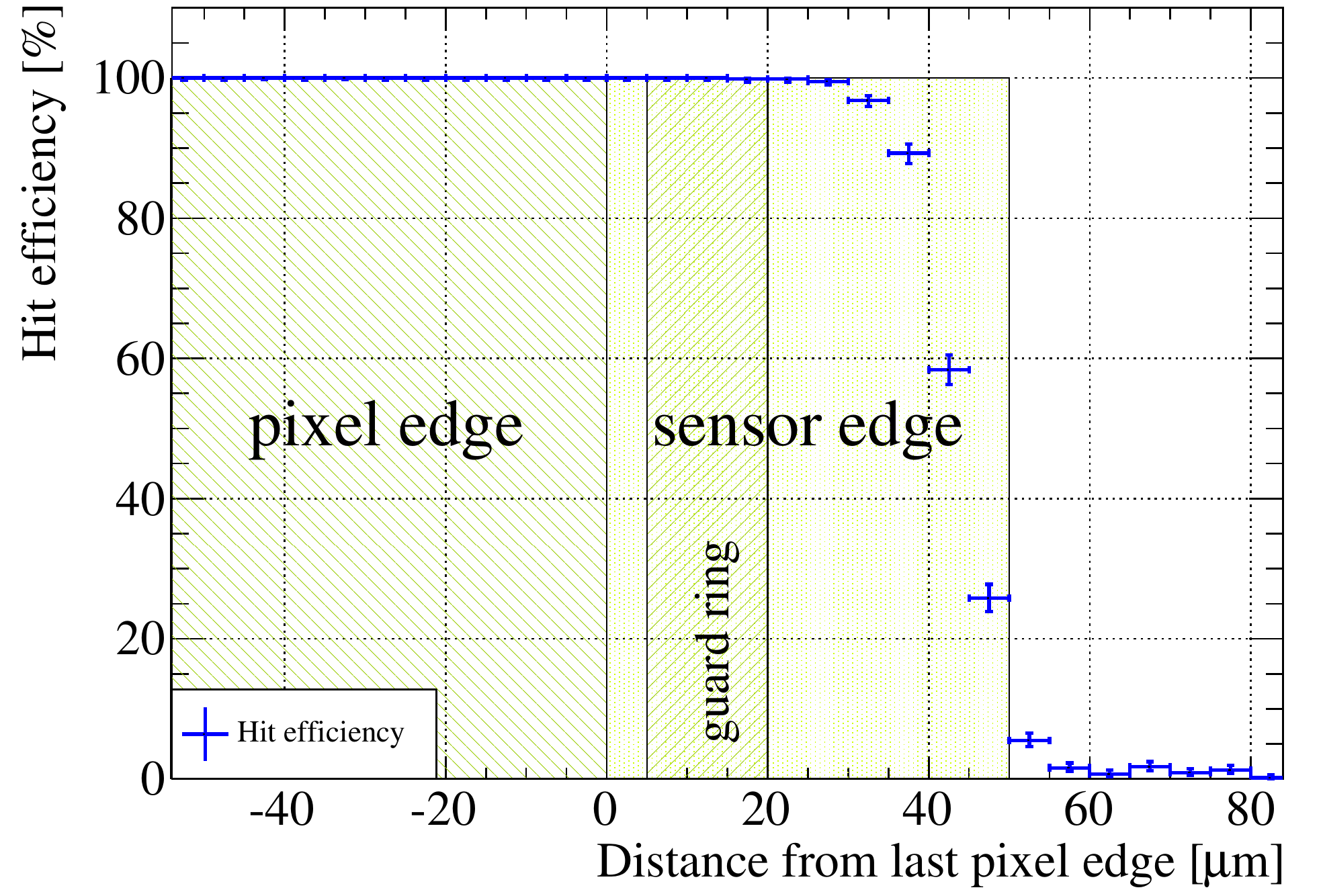}
	\label{fig:VTT-edge-50-b}
}
\caption{Characterization of the FE-I3 pixel module employing 100~\mum{} thin sensor with 50~\mum{} active edge design operated with a threshold of 1.5~ke: \protect\subref{fig:VTT-edge-50-a} shows the collected charge as a function of the bias voltage before and after irradiation to a fluence of 2$\times$10$^{15}$~\neqcm{}. After irradiation, the collected charge is measured before and after an annealing time of 90 hours at room temperature. After annealing, the results for the whole sensor, and for the edge pixels only, are compared. In~\protect\subref{fig:VTT-edge-50-b} the hit efficiency at the sensor edge is shown before irradiation.}
\label{fig:VTT-edge-50}
\end{figure*}

\paragraph{Edge analysis.} An analysis of the module performance in terms of charge collection and hit efficiency has been carried out for the slim and active edge designs. 
In figure~\ref{fig:VTT-edge-50-a} the collected charge is shown as a function of the bias voltage for the 50 \mum{} active edge design. Before irradiation the 100~\mum{} thin sensor shows the expected collected charge of almost 7~ke. At the fluence of 2$\times$10$^{15}$~\neqcm{} the measurement has been repeated after an annealing period of 90 hours at room temperature showing a significative increase of the collected charge up to almost 6~ke at 160~V, with perfectly compatible results between the edge pixels and the internal pixels. The hit efficiency for this module has been measured in a beam test at DESY before irradiation and the result is shown in figure~\ref{fig:VTT-edge-50-b}. The experimental setup has been optimized by reducing the distance of the two telescope arms thereby minimizing the impact of the multiple scattering on the telescope pointing resolution. An average hit efficiency of (87.4$\pm$0.7)\% has been measured in the last 50~\mum{} after the end of the pixel implant. This result is consistent with the same measurement performed at the CERN SpS with 120~GeV pions~\cite{slid-tsv-japan,iworid}.

\begin{figure*}[tbp] 
\centering
\subfigure[]{
	\includegraphics[width=.47\textwidth]{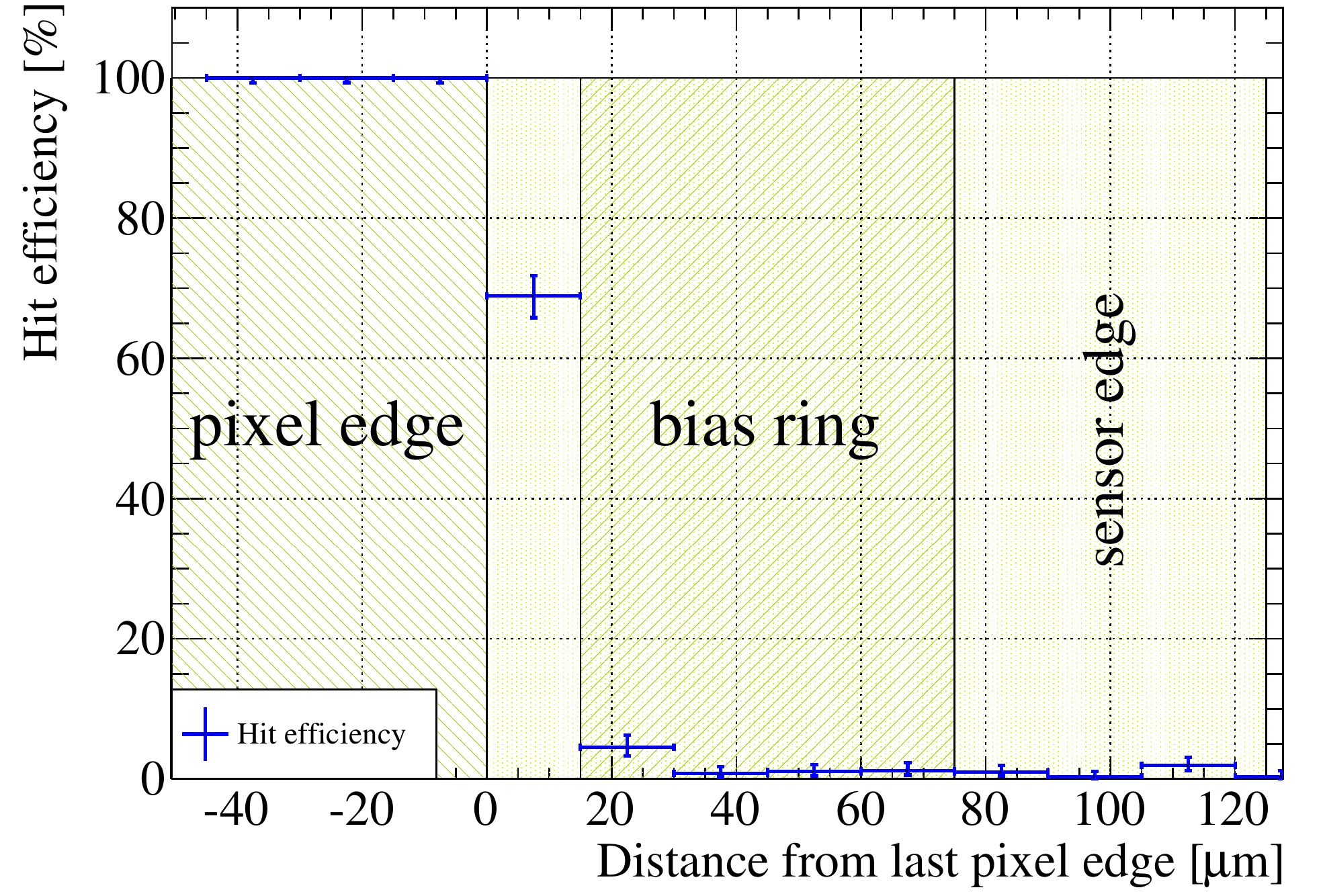}
	\label{fig:VTT-edge-125-a}
}
\subfigure[]{
	\includegraphics[width=.47\textwidth]{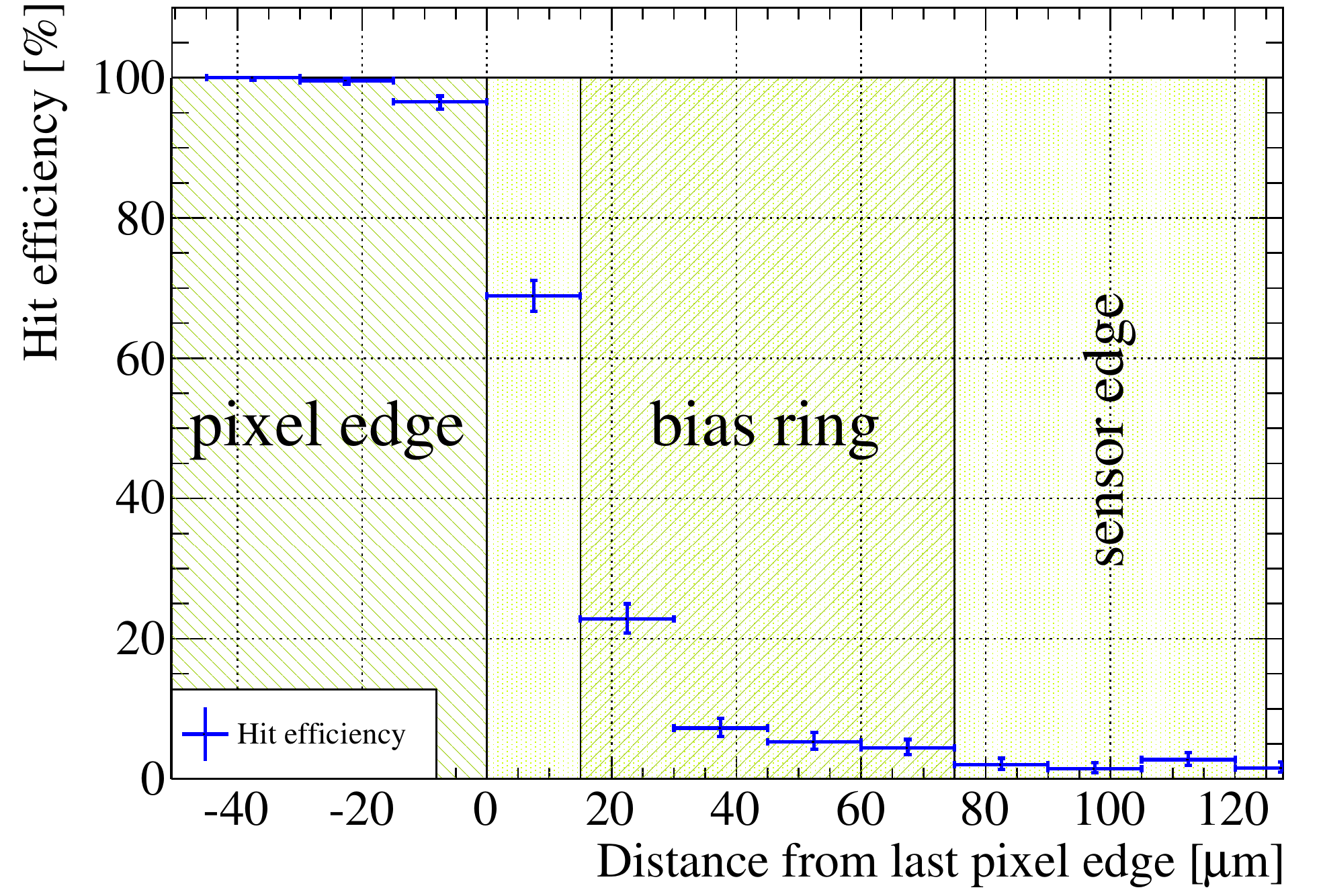}
	\label{fig:VTT-edge-125-b}
}
\caption{Hit efficiency at the edge of the 100~\mum{} thin FE-I3 pixel sensor with 125~\mum{} slim edge design operated with a threshold of 1.5~ke. The measurements have been performed at the CERN SpS \protect\subref{fig:VTT-edge-125-a} before irradiation and at DESY \protect\subref{fig:VTT-edge-125-b} after irradiation to a fluence of 5$\times$10$^{15}$~\neqcm{} at 300~V.}
\label{fig:VTT-edge-125}
\end{figure*}

Shown in figure~\ref{fig:VTT-edge-125} is the hit efficiency before and after irradiation at the edge of the FE-I3 module employing a 100~\mum{} thin sensor with d$_{\mathrm{edge}}=125$~\mum{}. In this edge design the BR is at ground potential but not read out, preventing the signal generated in the edge region to reach the active region. A residual hit efficiency of (69$\pm$3)\% is still observed between the end of the last pixel implant and the BR both before irradiation and after a fluence of 5$\times$10$^{15}$~\neqcm{}, where a value of (69$\pm$2)\% is reached increasing the bias voltage up to 300~V.

\begin{figure*}[tbp] 
\centering
	\subfigure[]{\includegraphics[width=.55\textwidth]{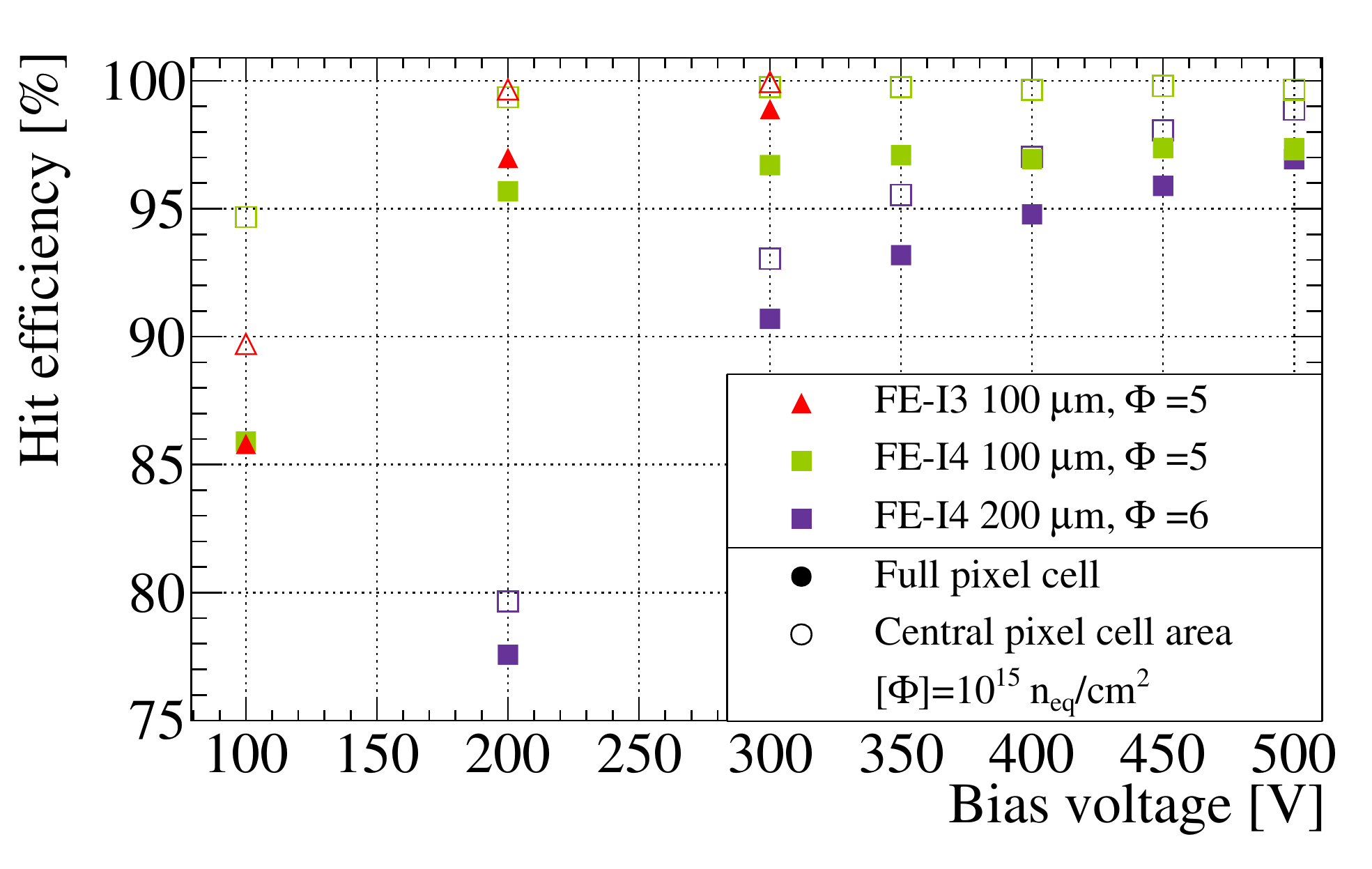}\label{fig:VTT-eff_a}}
	\subfigure[]{\includegraphics[width=.44\textwidth]{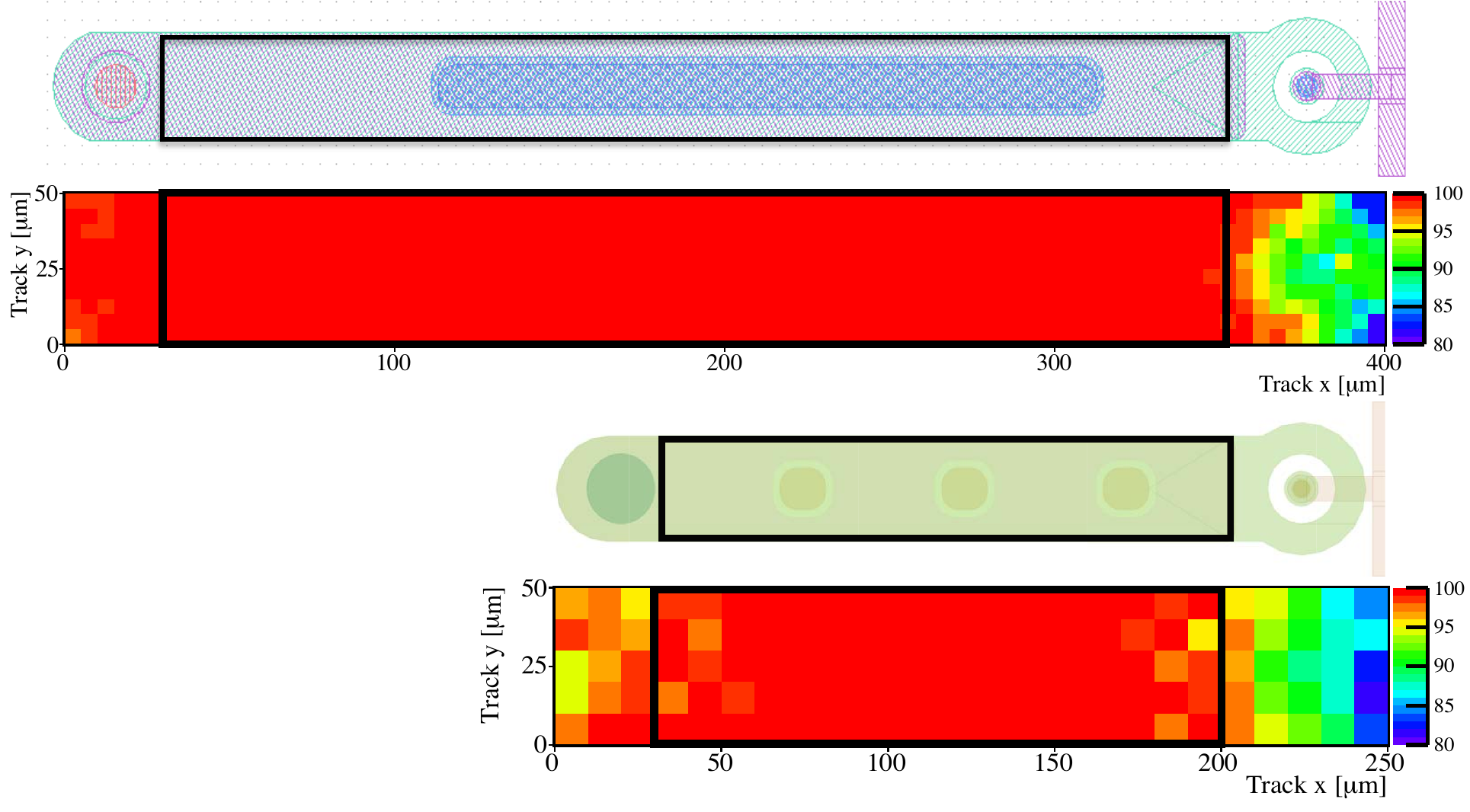}\label{fig:VTT-eff_b}}
\caption{Overview of the hit efficiency for VTT modules irradiated to a fluence of 5-6$\times$10$^{15}$~\neqcm{}. In~\protect\subref{fig:VTT-eff_a} the hit efficiency as a function of the bias voltage is compared between 100~\mum{} and 200~\mum{} thick sensors. The full symbols represent the hit efficiency over the full module. The open symbols are the hit efficiency calculated for the center of the pixel cell as represented by the black box in~\protect\subref{fig:VTT-eff_b} for the FE-I3 cell 50$\times$400~\mum{}$^2$ (top) and the FE-I4 cell 50$\times$250~\mum{}$^2$ (bottom). In~\protect\subref{fig:VTT-eff_b} the hit efficiency measured at the highest voltage for the two modules with 100~\mum{} thin sensor is projected onto a the single pixel cell. The 100~\mum{} thin FE-I3 and the two 100 and 200~\mum{} thick FE-I4 assemblies have been tuned to a threshold of 1.5, 2.3 and 1.6~ke, respectively.}
\label{fig:VTT-eff}
\end{figure*}

Figure~\ref{fig:VTT-eff} shows the hit efficiency measured at DESY of the VTT modules irradiated to a fluence of 5-6$\times$10$^{15}$~\neqcm{}. 
The FE-I4 module with 200~\mum{} thick sensor reaches a 97\% hit efficiency at 500~V while the same module type with 100~\mum{} thick sensor starts to saturate to this value already at 300~V. As shown in figure~\ref{fig:VTT-eff_b} the region with reduced efficiency after irradiation corresponds to the punch through area that is not directly connected to the pixel implant. Because of the smaller fraction of area that this structure occupies for the the FE-I3 module, the latter shows a higher hit efficiency than the FE-I4 module with the same sensor thickness. Restricting the analysis to the central area of the pixel cell, by exclusion of the punch through region and also of the area of charge sharing between four adjacent pixels, the FE-I3 and FE-I4 sensors irradiated at the same fluence show consistent results with a saturation of the hit efficiency around 99.7\% at 300~V.

\subsection{The n-in-p pixel production at CiS}\label{sec:CiS}
\begin{figure*}[tbp] 
\centering
\subfigure[]{\includegraphics[width=0.34\textwidth]{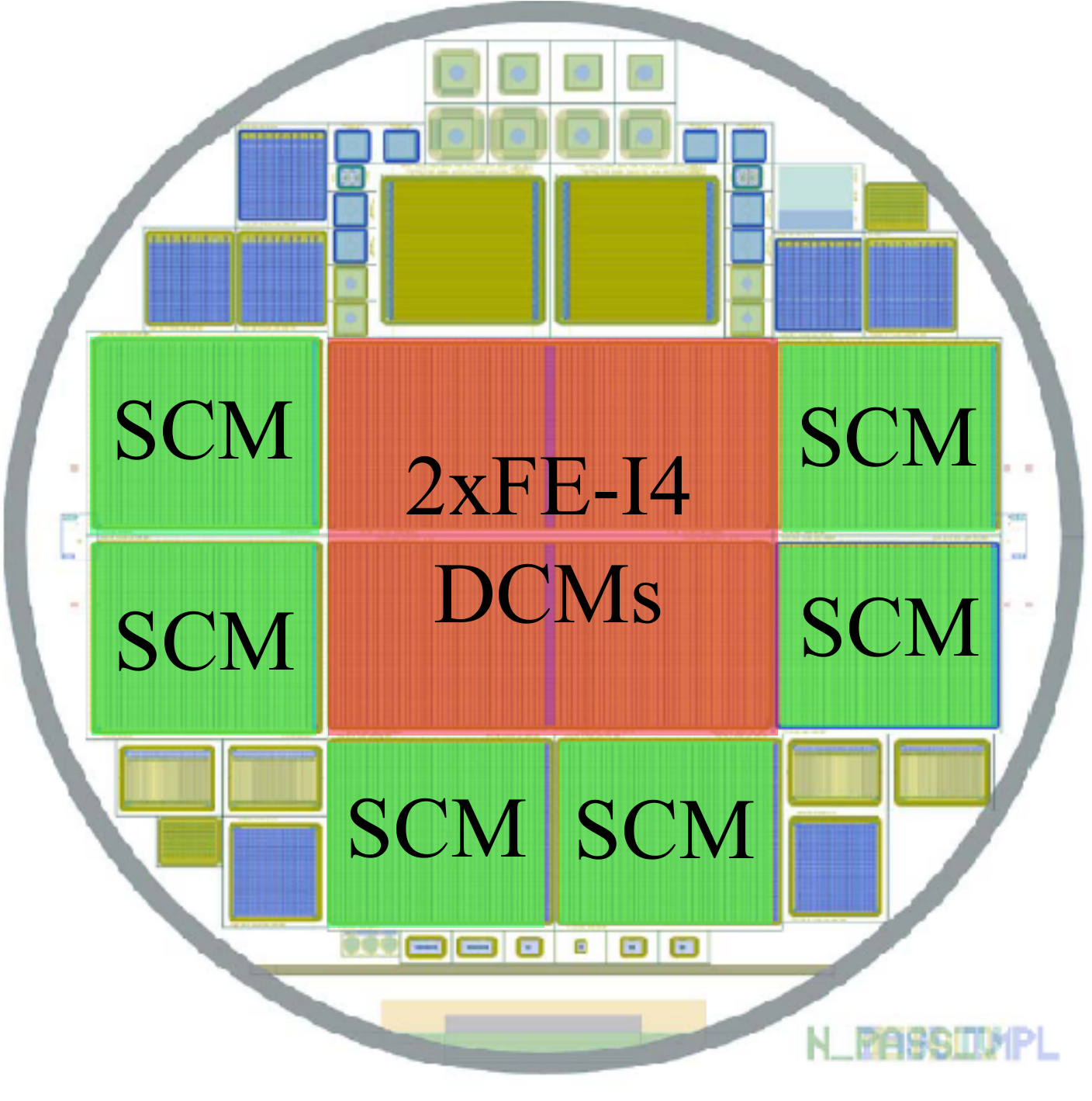}\label{fig:IV_CiS_a}}
\subfigure[]{\includegraphics[width=0.61\textwidth]{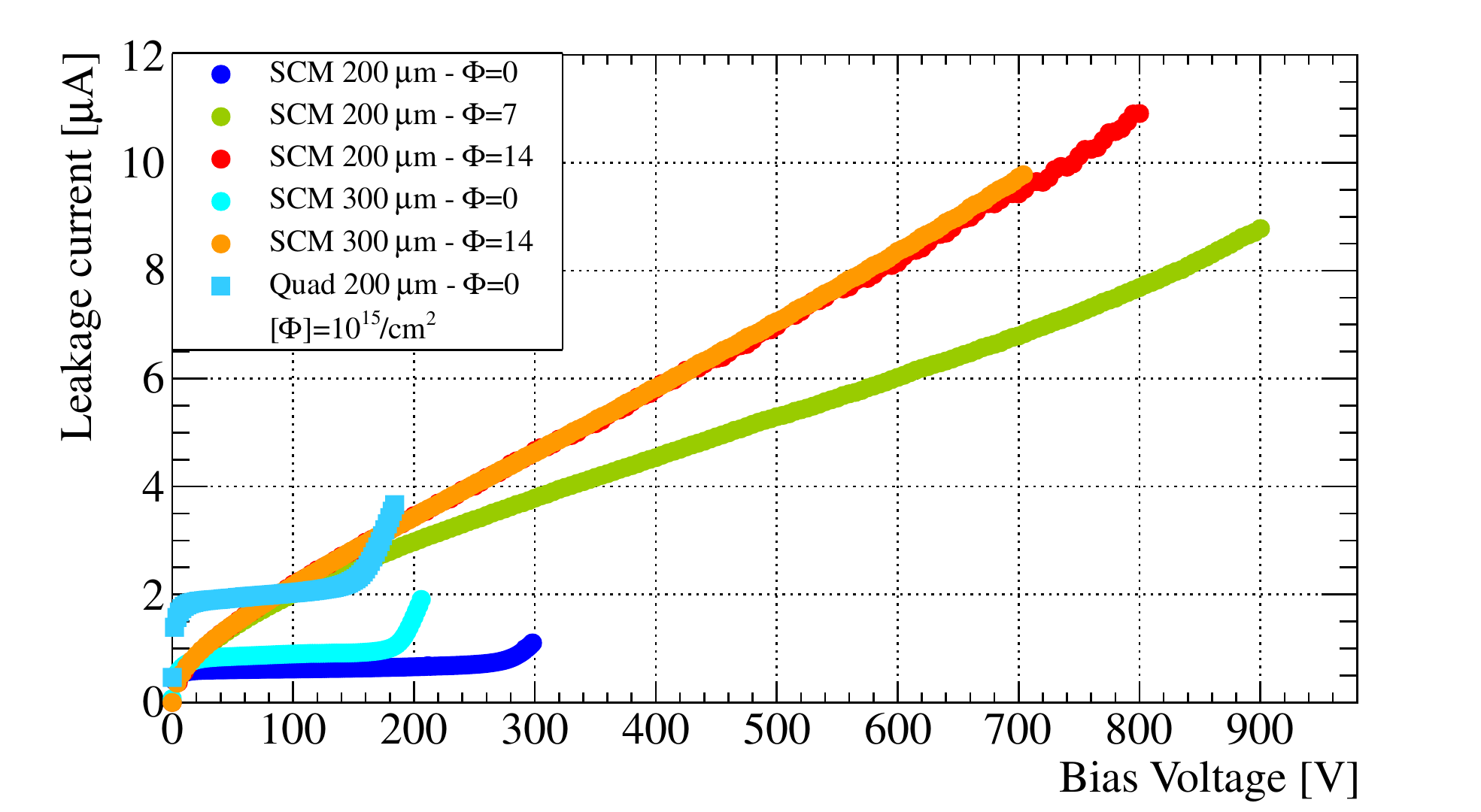}\label{fig:IV_CIS_b}}
\caption{The n-in-p silicon pixel production at CiS. The wafer layout in~\protect\subref{fig:IV_CiS_a} states the location of the Single Chip (SCM), and the Double Chip Modules (DCM) to be interconnected as a quad module. In~\protect\subref{fig:IV_CIS_b} the IV curves after interconnection to ATLAS FE-I4 chips measured before and after irradiation are shown.}
\label{fig:IV_CIS}
\end{figure*}

Planar n-in-p pixel sensors have been produced at CiS on 4 inch wafers of p-type Float Zone (FZ) silicon with a resistivity of 15~k$\Omega\,$cm and an active thicknesses of 200 or 300~\mum{}. The sensors are designed with 10 GRs and one BR structure, and exhibit a distance of 450~\mum{} from the last pixel implant to the sensor edge. Homogeneous p-spray with Boron ions on the front side has been used for implant isolation and the sensor surface is covered with a 3~\mum{} layer of BenzoCycloButene (BCB) to prevent sparks between the sensor edges at high voltage and the chip at ground potential. This production includes Single Chip Module (SCM) sensors that have been interconnected to FE-I4 chips, and 2$\times$Double Chip Module (DCM) sensors diced as a single sensor piece and interconnected to four FE-I4 chips. The latter design follows the one planned for the outer layers of the ATLAS pixel detector at HL-LHC and in the following it will be referred to as quad module. This production is part of a collaboration between the Max-Planck-Institut f\"ur Physik and the Universities of G\"ottingen and Bonn.
Shown in figure~\ref{fig:IV_CIS_b} are the IV curves of the tested modules before and after irradiation in Los Alamos up to a fluence of 14$\times$10$^{15}$~\neqcm{}. Before irradiation, all the structures show a breakdown voltage well above the full depletion that occurs around 30~V for the 200~\mum{} thick sensors and around 60~V for the 300~\mum{} thick sensors. No breakdown is observed for the irradiated sensors up to 800~V.

\subsubsection{Single chip module characterization}\label{sec:SCM}
\begin{figure*}[tbp] 
\centering
\subfigure[]{
	\includegraphics[width=.36\textwidth]{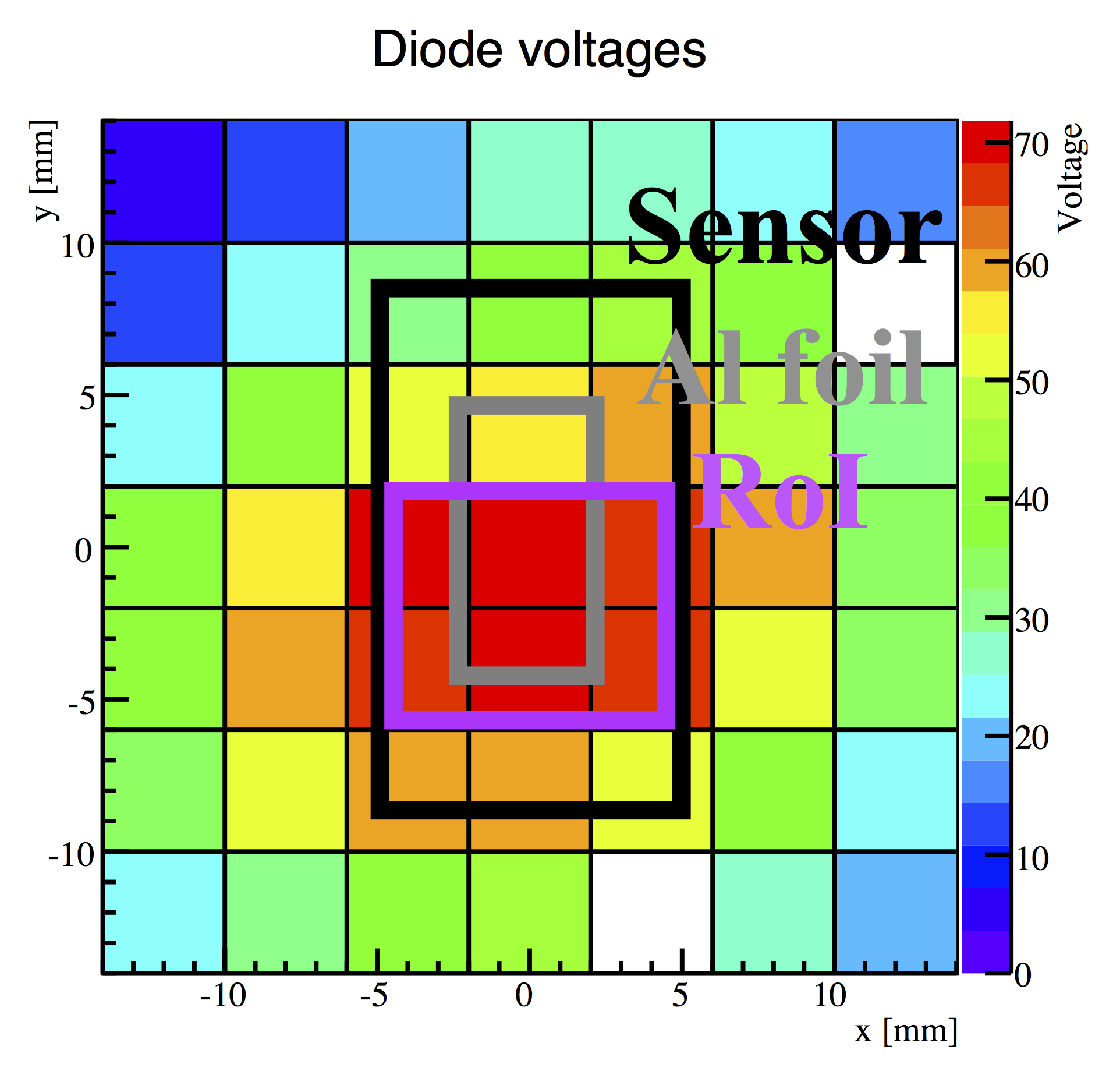}
	\label{fig:diode_and_cc_a}
}
\subfigure[]{
	\includegraphics[width=.46\textwidth]{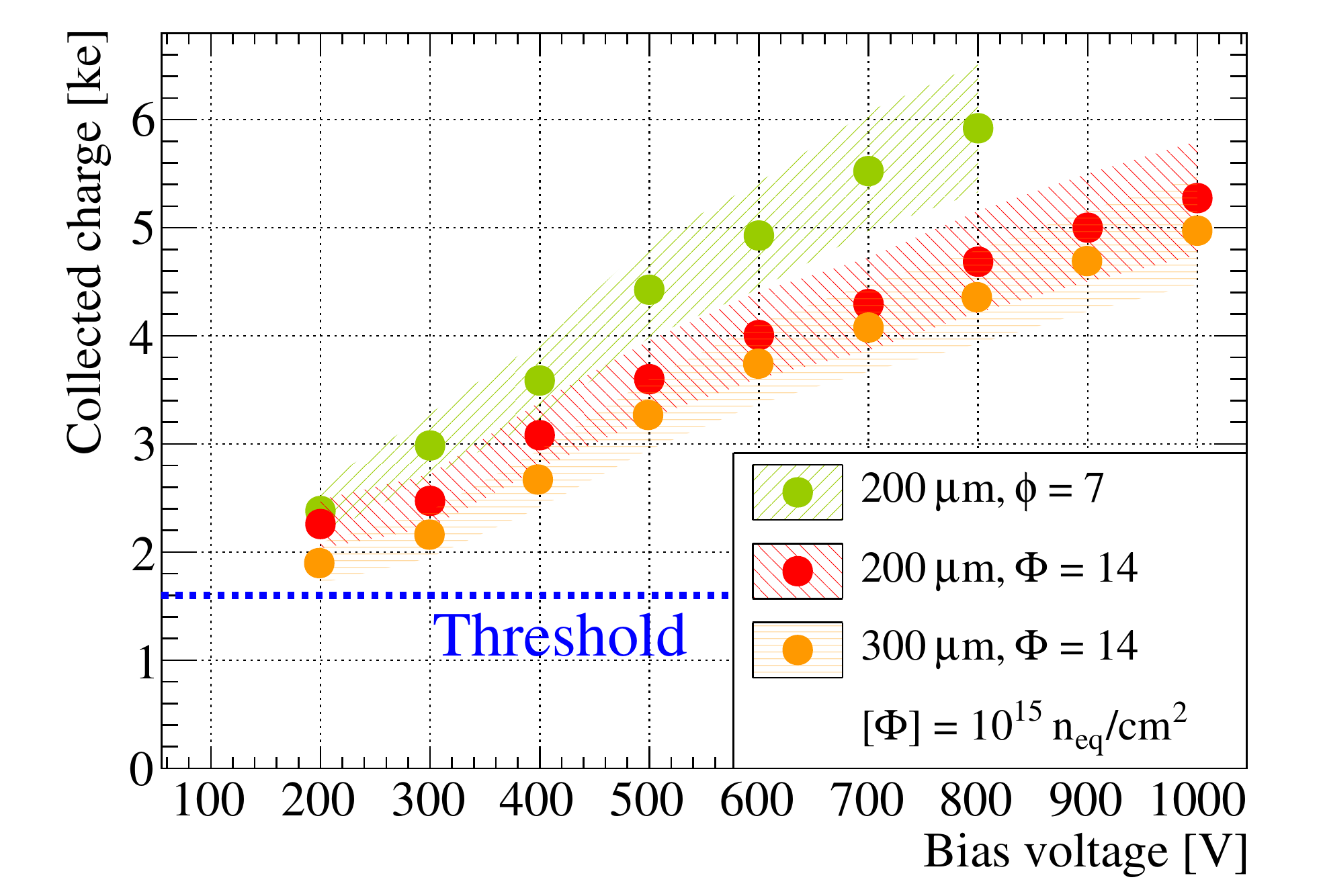}
	\label{fig:diode_and_cc_b}
	}
\caption{Figure~\protect\subref{fig:diode_and_cc_a} shows the diode measurement for the irradiation of the CiS modules in Los Alamos including the position on the sensor as well as the aluminum foil and the RoI used for the analysis. The module and the aluminum foil are tilted by 30$^{\circ}$ with respect to the proton beam and the diode to increase the uniformity of the irradiation. \protect\subref{fig:diode_and_cc_b} shows the collected charge as a function of the bias voltage calculated inside the RoI for different thicknesses and irradiation fluences.}
\label{fig:diode_and_cc}
\end{figure*}

The SCMs have been irradiated in Los Alamos with an 800~MeV proton beam to fluences of 7 and 14$\times$10$^{15}$~\neqcm{}.
The received fluence is not uniform over the module surface due to the small size of the proton beam with respect to the sensor dimensions. 
To obtain a characterization of the modules that corresponds to a specific fluence, the analysis has been performed inside a Region of Interest (RoI) corresponding to the center of the beam spot. The RoI has been defined rescaling the average fluence measured by a 1$\times$1~cm$^{2}$ aluminum foil into a more uniform region of the beam spot. This region and its average fluence has been extrapolated from the diode array used at the Los Alamos facility to monitor the beam profile~\cite{los-alamos}. The approximate position of the aluminum foil, the diode and the modules is sketched in figure~\ref{fig:diode_and_cc_a}. 
A 10\% uncertainty on the fluence has been estimated considering the uncertainty on the aluminum foil measurements and on the setup positions.

\paragraph{Charge collection.} The collected charge of the irradiated SCMs has been measured inside the RoI as a function of the bias voltage. Results obtained for the different thicknesses and irradiation fluences are compared in figure~\ref{fig:diode_and_cc_b}. After irradiation, the collected charge does not saturate up to the maximum applied voltage of 1000~V. At the highest fluence of 14$\times$10$^{15}$~\neqcm{} the collected charge of the 200~\mum{} thick sensors is slightly higher than for the ones 300~\mum{} thick, but still compatible within the estimated uncertanties.

\begin{figure*}[tbp] 
\centering
\subfigure[]{\includegraphics[width=0.49\textwidth]{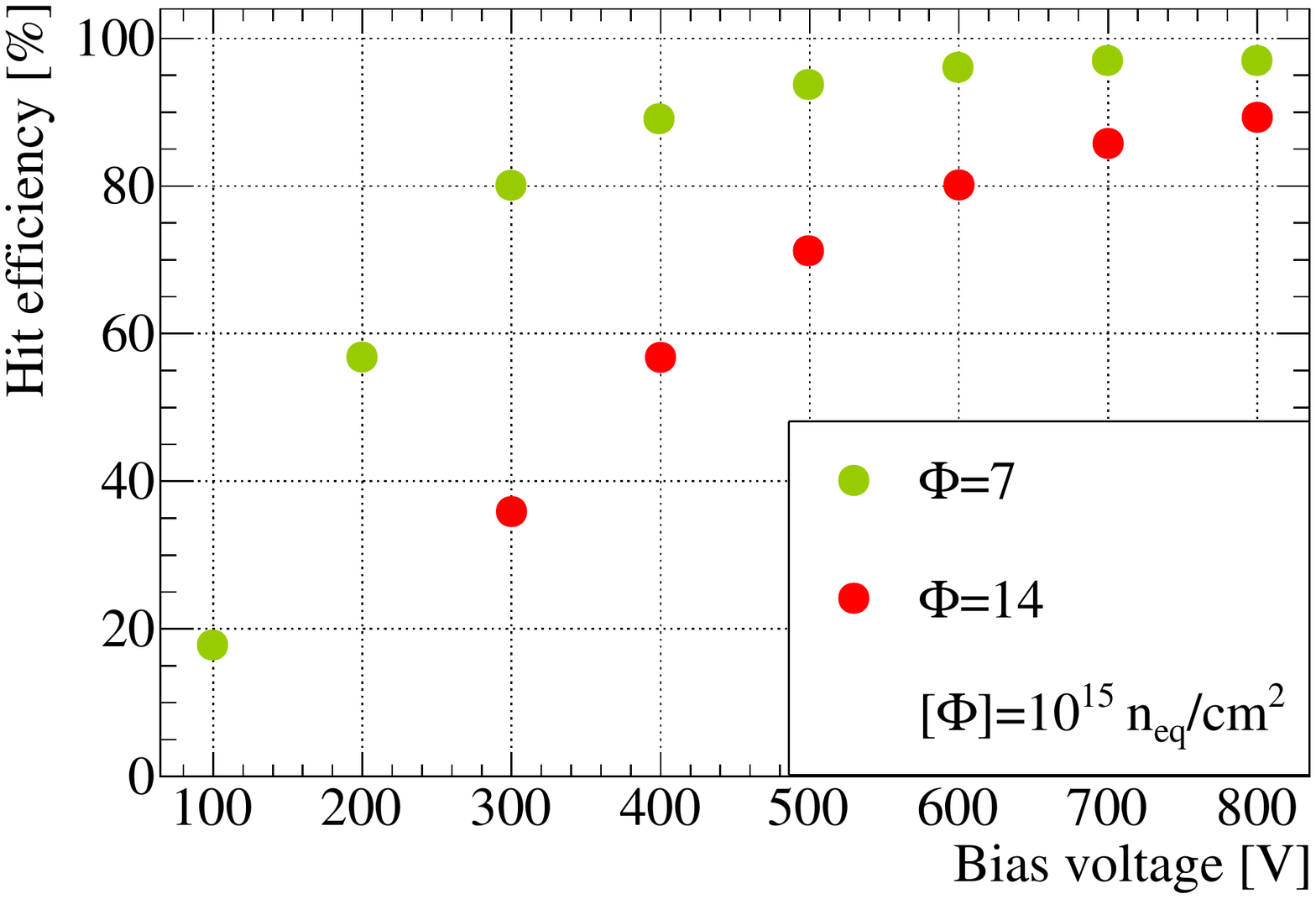}\label{fig:CiS_eff_a}}
\subfigure[]{\includegraphics[width=0.325\textwidth]{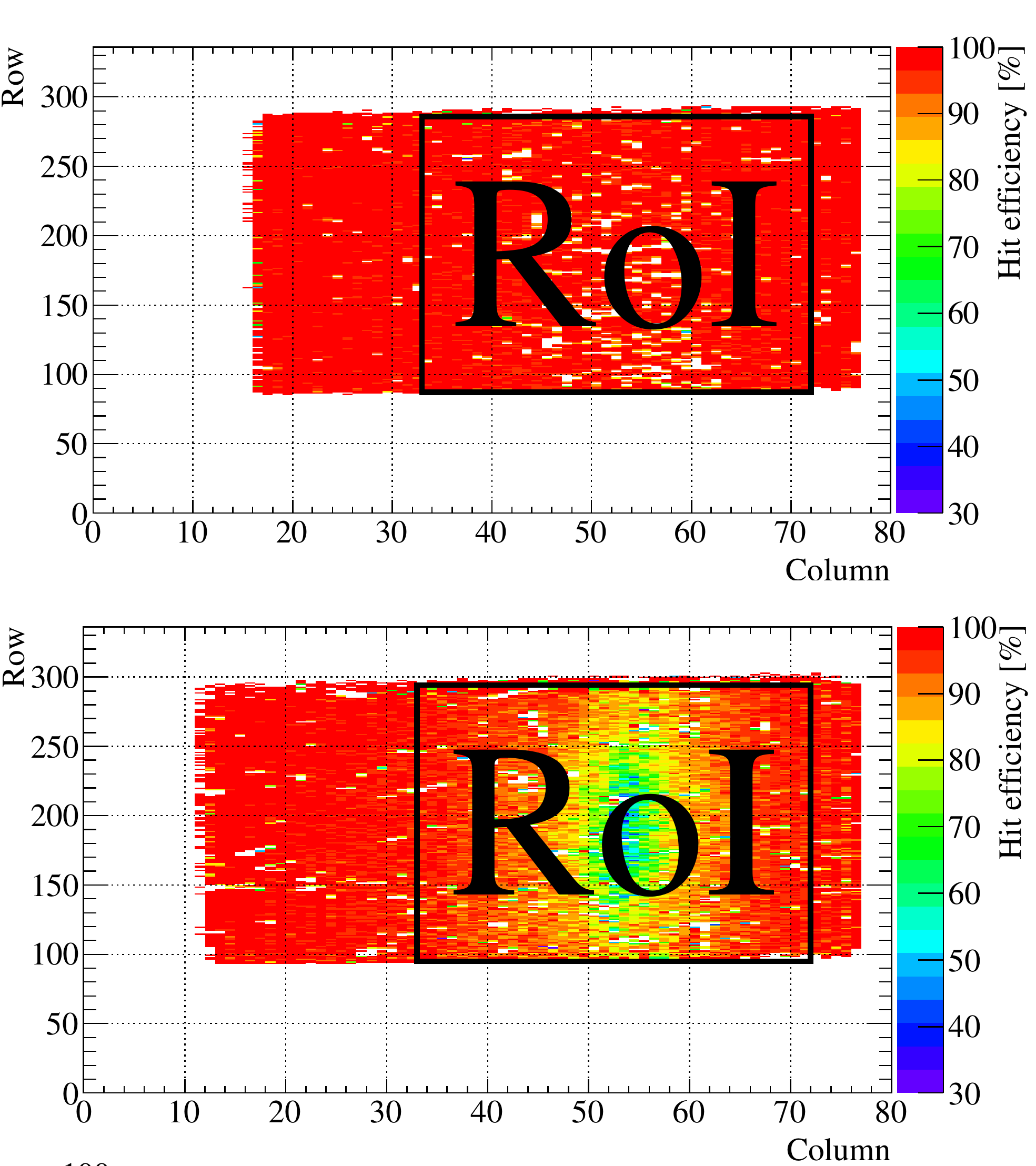}\label{fig:CiS_eff_b}}
\caption{In~\protect\subref{fig:CiS_eff_a} the hit efficiency as a function of the bias voltage is shown for the 200~\mum{} thick CiS modules irradiated in Los Alamos operated with a threshold of 1.6~ke. The RoI considered for the analysis is highlighted in~\protect\subref{fig:CiS_eff_b}, showing the distribution of the hit efficiency across the sensor surface for the CiS module irradiated to 7 (top) and 14$\times$10$^{15}$~\neqcm{} (bottom). This results are obtained at a bias voltage of 800~V.}
\label{fig:CiS_eff}
\end{figure*}

\paragraph{Hit efficiency.} The hit efficiency as a function of the bias voltage has been measured with a perpendicular incident beam at DESY for the SCMs with 200~\mum{} thick sensors after irradiation. The results are compared in figure~\ref{fig:CiS_eff_a}. At the fluence of 7$\times$10$^{15}$~\neqcm{} an hit efficiency of 97.2\% is obtained increasing the bias voltage to 800 V. At the same bias voltage, the module irradiated to a fluence of 14$\times$10$^{15}$~\neqcm{} shows a hit efficiency of only 89.4\%. Shown in figure~\ref{fig:CiS_eff_b} are the efficiency maps for the two measurement at the highest voltage. In the map of the module irradiated at a fluence of 14$\times$10$^{15}$~\neqcm{}, the effect of the beam spot from the not uniform irradiation is clearly visible.

\subsubsection{Four chip module characterization}\label{sec:SCM}
The quad modules employ a 200~\mum{} thick sensor interconnected to four FE-I4 chips that have been thinned down to 150~\mum{} at IZM. A glass substrate has been attached to the chip backside to avoid the bending of the corners after thinning, that could lead to a decrease of the bump-bonding efficiency~\cite{izm}.
The modules have an inactive area in the 1.6 mm gap between the rows of the two double sensors, where the GR and BR structures are located. Moreover, the two pixel columns in the middle of each double sensor are 450~\mum{} long to allow the necessary space for the placement of the two neighboring chips. The full modules are wire bonded to detector boards designed by the University of Liverpool.

\begin{figure*}[tbp] 
\centering
\subfigure[]{
	\includegraphics[width=0.46\textwidth]{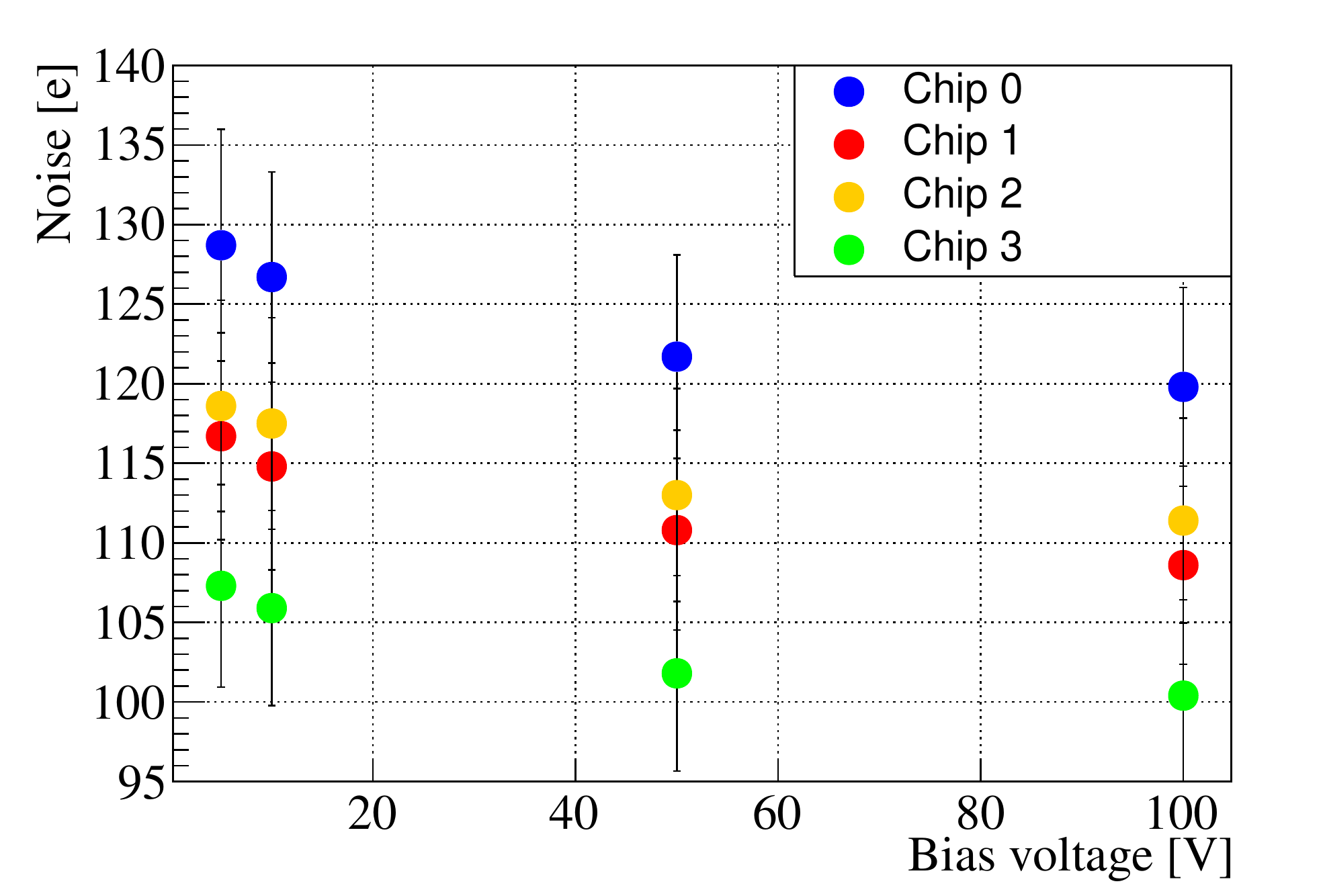}\label{fig:Quad-cc_a}
}
\subfigure[]{
	\includegraphics[width=0.46\textwidth]{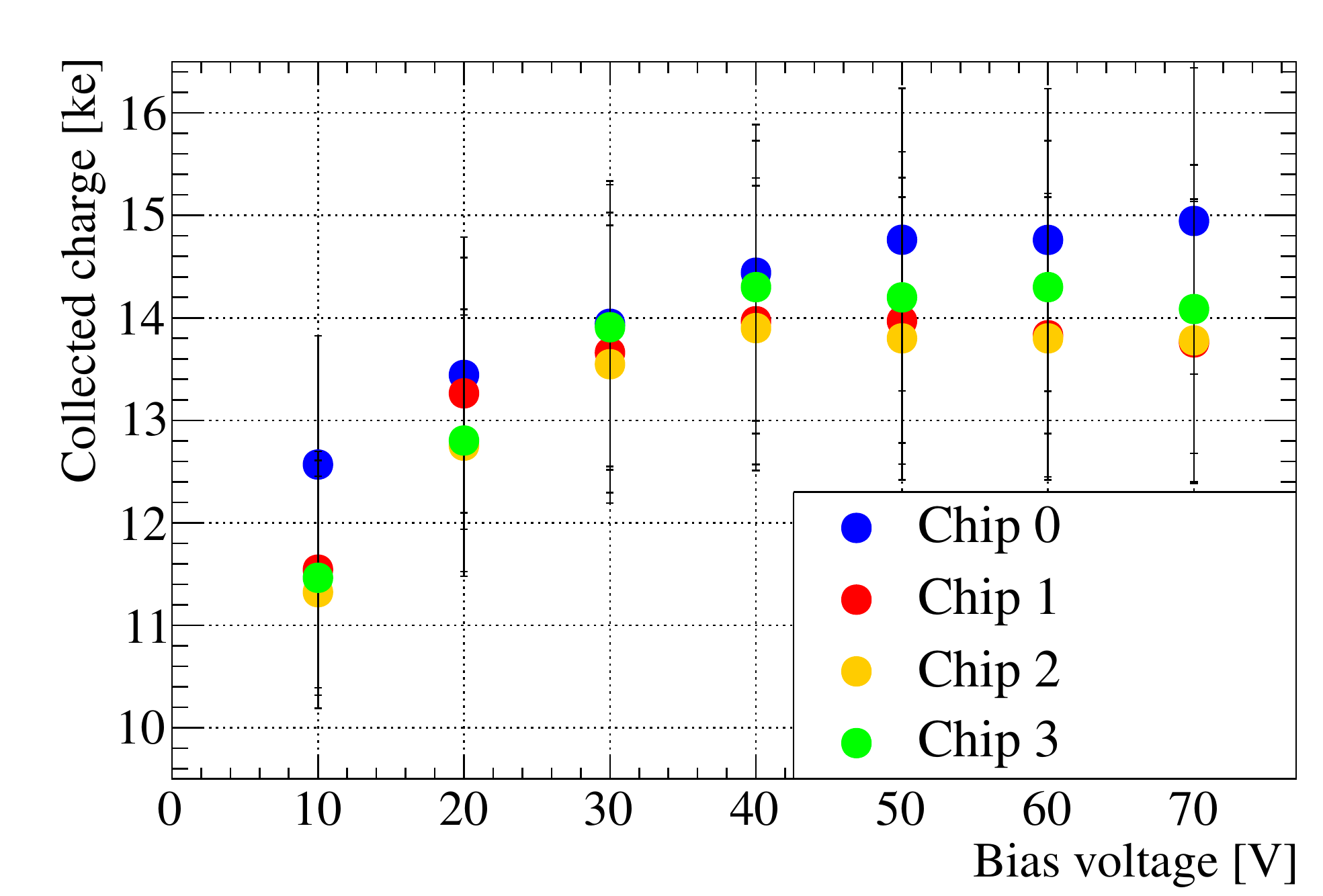}\label{fig:Quad-cc_b}
}
\caption{In figure~\protect\subref{fig:CiS_eff_a} the average noise values are shown for each of the four chips of the quad module as a function of the bias voltage. The module was tuned to 7~ToT at 14~ke with a threshold of 3~ke. In~\protect\subref{fig:CiS_eff_b} their collected charge as a function of the bias voltage are given.}
\end{figure*}

\paragraph{Charge collection.} The collected charge as a function of the bias voltage has been measured using the USBPix system and a ``burn-in'' adapter card~\cite{usbpix} that is able to read out  four chips in parallel. The module has been tuned with a threshold of 3~ke. Results in figure~\ref{fig:Quad-cc_b} show that the collected charge is consistent with the expected value of almost 14~ke for a 200~\mum{} thick silicon sensor when the applied bias voltage is above the full depletion voltage. Shown in figure~\ref{fig:Quad-cc_a} is the noise as a function of the bias voltage for the four chips for the same tuning. Above the full depletion voltage the noise is stable with values between 100 and 120~e, and it is comparable to the typical noise level of a SCM from the same production.

\begin{figure*}[tbp] 
\centering
\subfigure[]{
	\includegraphics[width=0.48\textwidth]{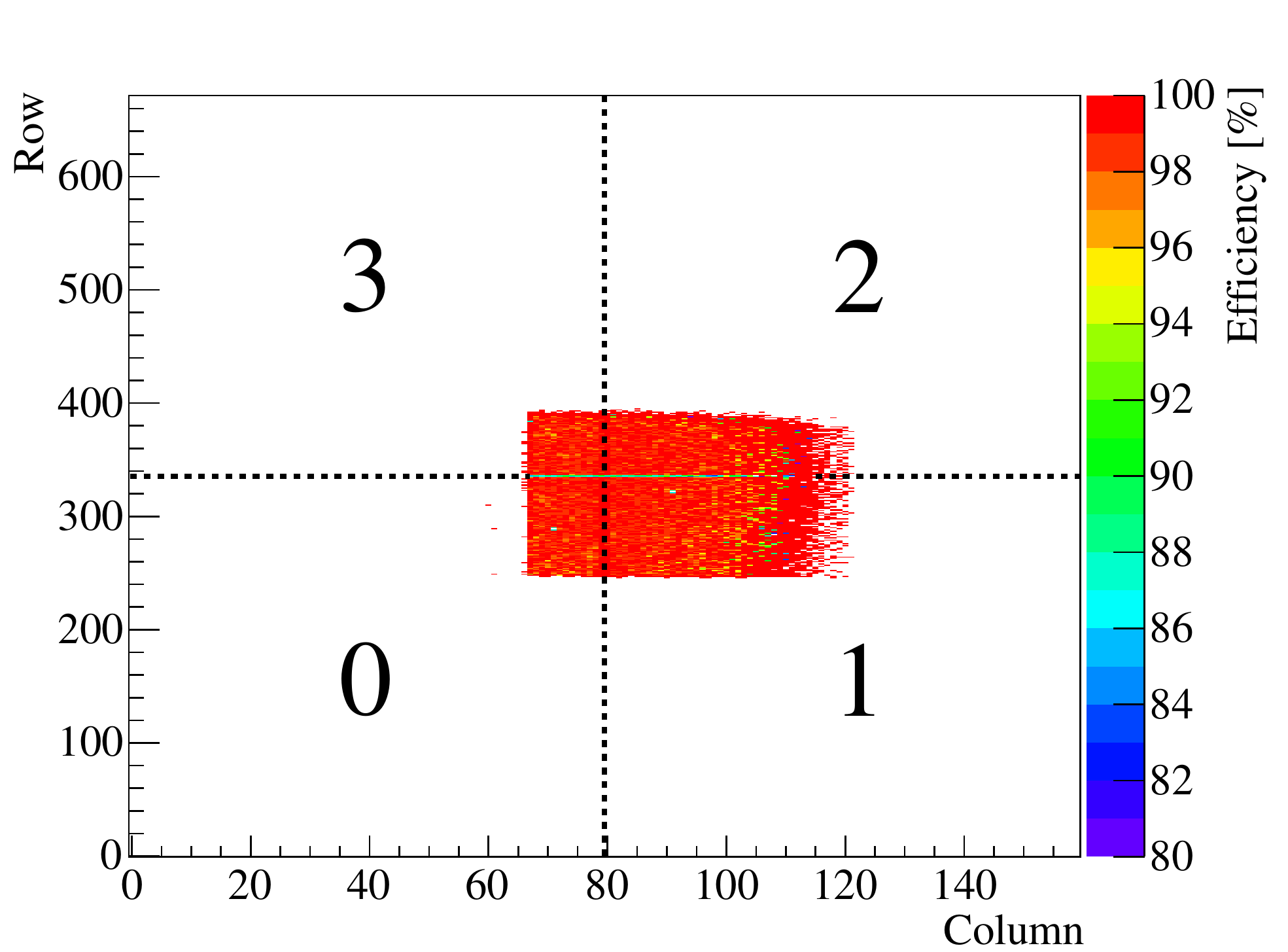}\label{fig:Quad-eff_a}
}
\subfigure[]{
	\includegraphics[width=0.48\textwidth]{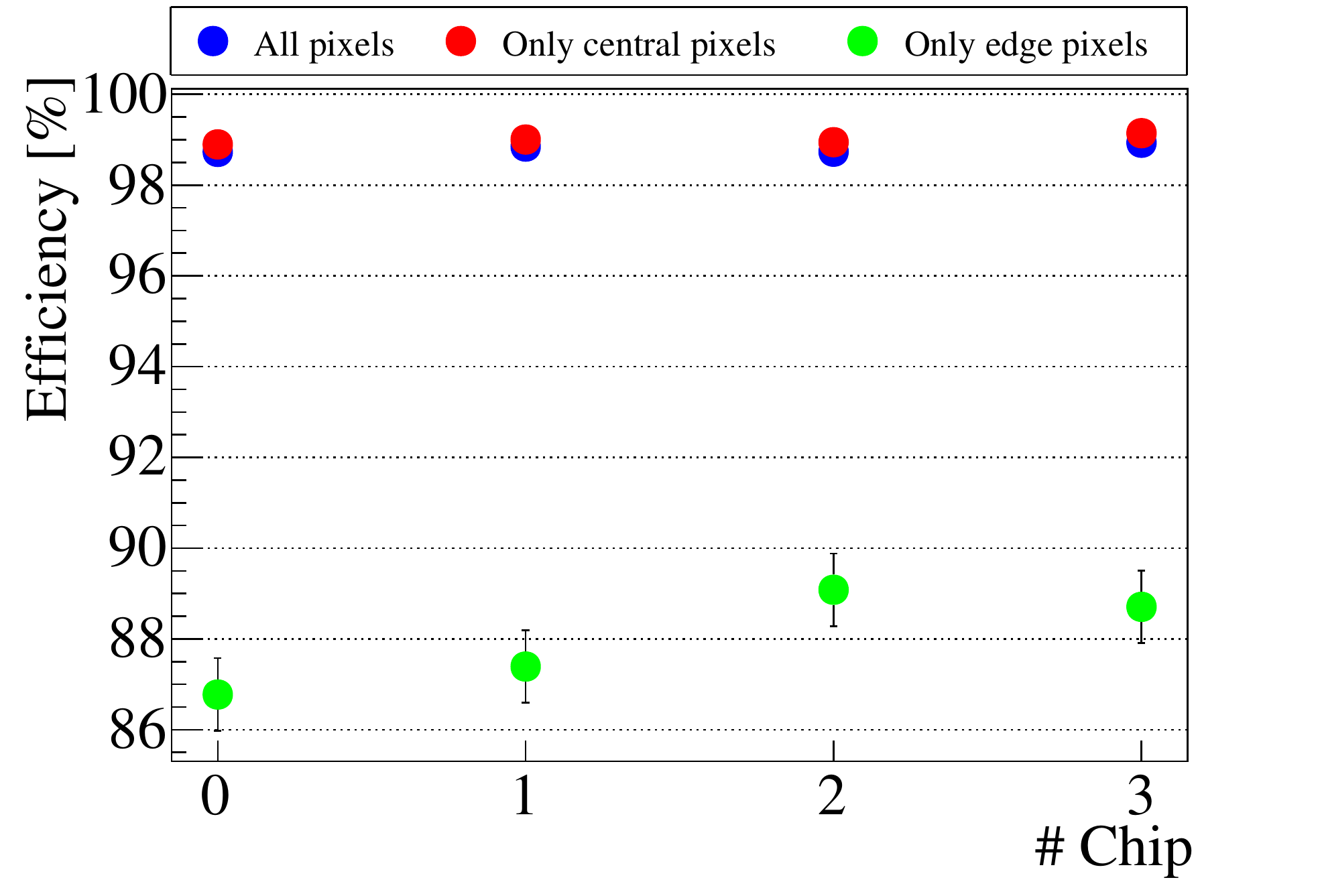}\label{fig:Quad-eff_b}
}
\caption{Hit efficiency of the quad module: \protect\subref{fig:Quad-eff_a} shows the distribution over the full module surface. Each of the chip couple 0-1 and 2-3 is interconnected to a different double sensor. \protect\subref{fig:Quad-eff_b} shows the hit efficiency chip by chip for all pixel cells (blue dots), for the pixel cells at the edge between the two double sensors (green dots), and for all pixel cells but the ones at the edge between the two double sensors (red dots).}
\label{fig:Quad-eff}
\end{figure*}

\paragraph{Hit efficiency.} The hit efficiency for perpendicular incident particles obtained at DESY is shown in figure~\ref{fig:Quad-eff}. The threshold has been tuned to 3~ke and the chips are simultaneously read out with the RCE readout system~\cite{rce}. The beam has been centered in the module to study the efficiency of the critical areas of the assembly where all four chips are involved in the data taking. The full module shows an overall efficiency of 99\%. A lower hit efficiency of the pixel rows at the edge between the two double chip sensors is observed due to multiple scattering that degrades the telescope pointing resolution and leads to the inclusion in the efficiency calculation of tracks that are crossing the inactive area between the two double sensors.

\subsection{Thickness comparison}\label{sec:thickness}
\begin{figure*}[tbp] 
\centering
\subfigure[]{ \includegraphics[width=.47\textwidth]{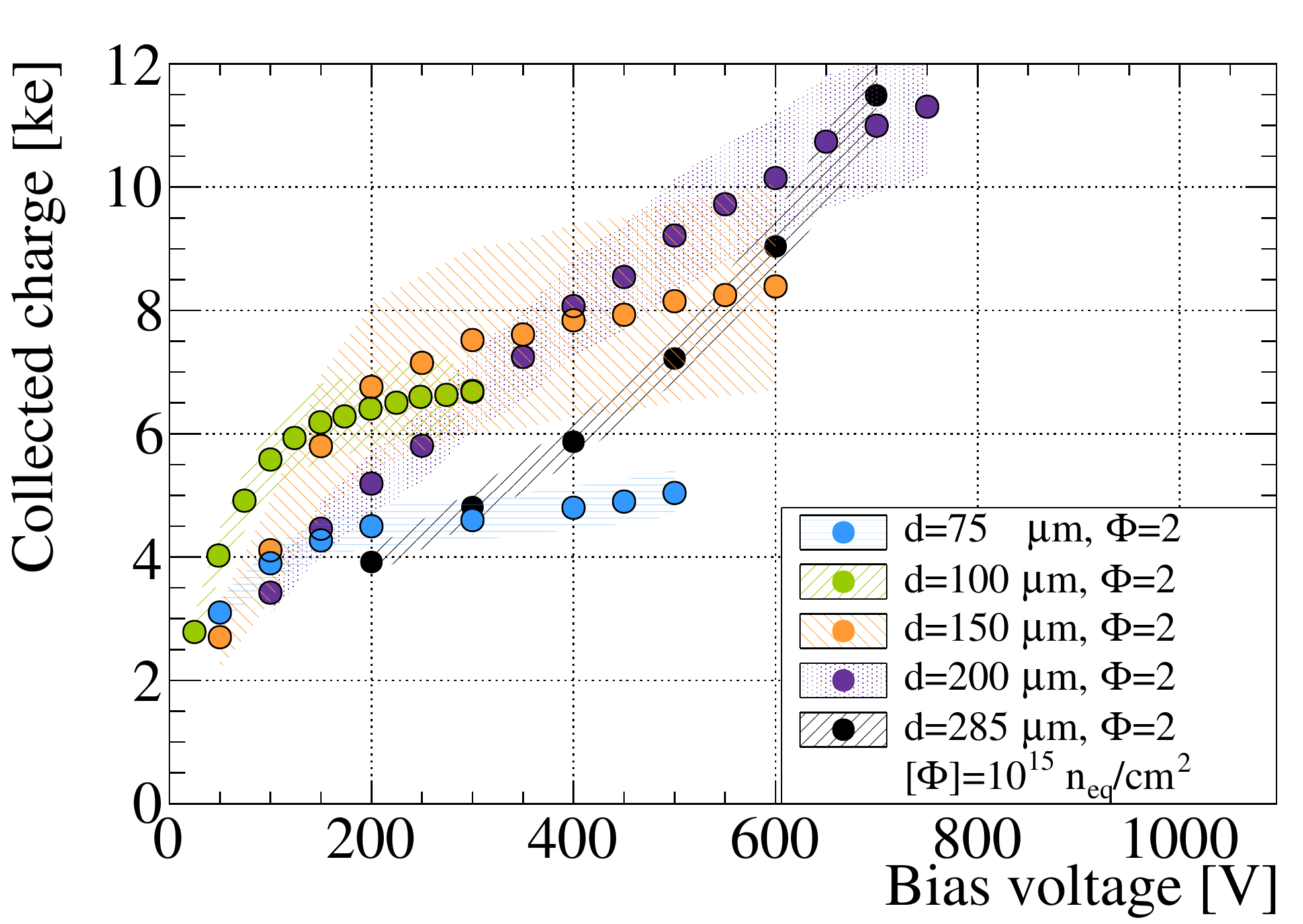} \label{fig:thick-comp-2e15} }
\subfigure[]{ \includegraphics[width=.477\textwidth]{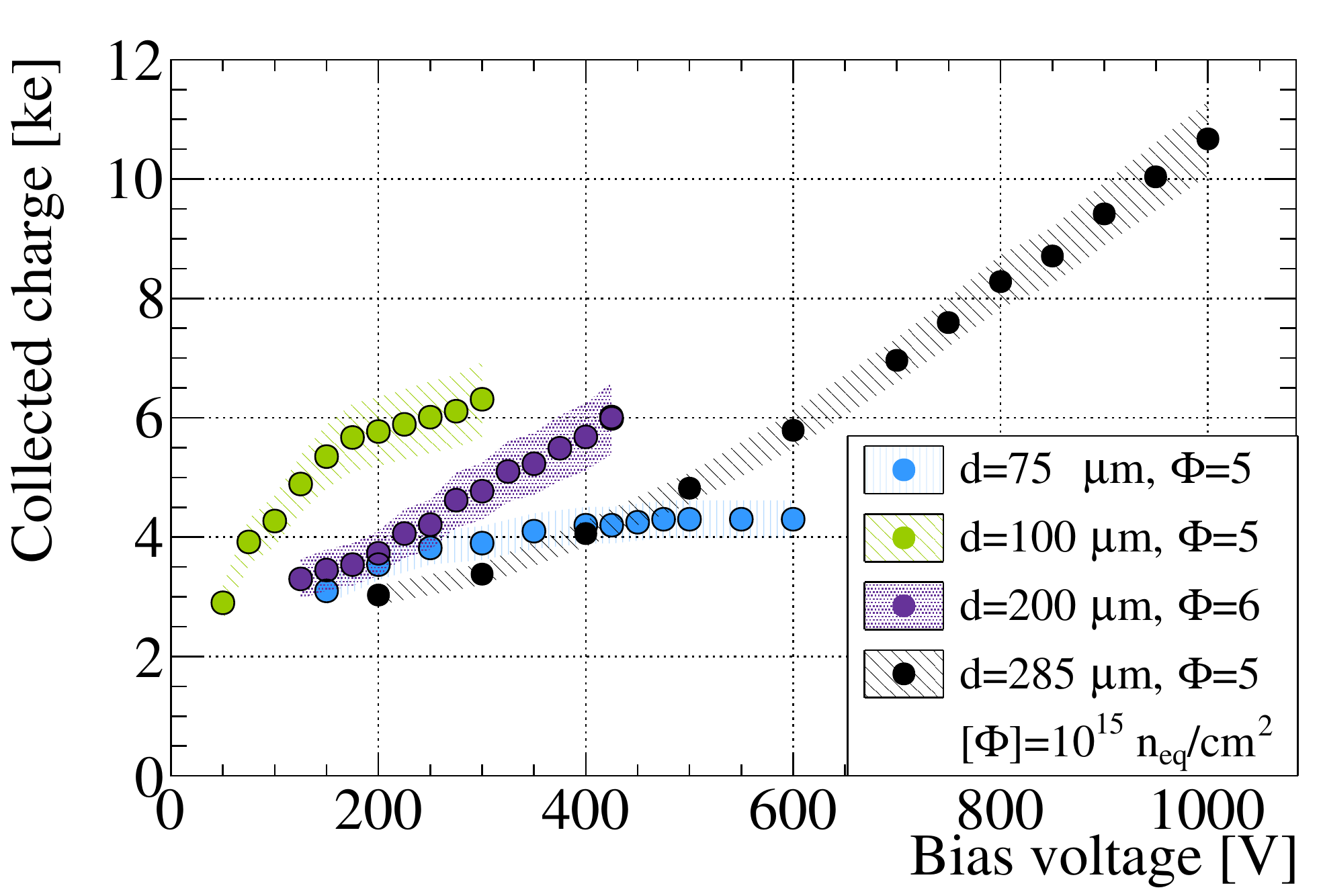} \label{fig:thick-comp-5e15} }
\caption[]{Overview of the charge collection for irradiated pixel sensors of different thicknesses. The results after a fluence of 2$\times$10$^{15}$ \neqcm{} and 5-6$\times$10$^{15}$~\neqcm{} are shown in~\protect\subref{fig:thick-comp-2e15} and~\protect\subref{fig:thick-comp-5e15}, respectively.}
\label{fig:thick-comp}
\end{figure*}

The charge collection for different sensor thicknesses are compared at the fluences of 2 and 5-6$\times$10$^{15}$~\neqcm{} in figure~\ref{fig:thick-comp}. After irradiation, the collected charge of the thin sensors ($\leq150$~\mum{}) starts to saturate around 200~V showing the highest collected charge for moderate voltages between 200 and 300~V, while the thicker sensors are still under depleted up to 700~V after a fluence of 2$\times$10$^{15}$~\neqcm{} and even at 1000~V after a fluence of 5$\times$10$^{15}$~\neqcm{}. 

\section{Conclusions}\label{sec:conclusions}
From the characterization of the VTT active edge modules before irradiation, the 100~\mum{} thin sensors with 50~\mum{} active edge are efficient even outside the pixel area up to the sensor border. After irradiation to the fluence of 2$\times$10$^{15}$~\neqcm{} the same charge collection efficiency of 84\% at 160~V has been measured for the inner pixels and the edge pixels. The 125~\mum{} slim edge design is active up to the BR both before and after irradiation to 5$\times$10$^{15}$~\neqcm{}. The irradiated 100~\mum{} thin sensors show an early saturation of the collected charge between 200 and 300~V. Their global hit efficiency reach about 97\% already at 300~V while the 200~\mum{} thick sensors need at least 500~V to achieve similar values. The punch through region has been identified as the main inefficiency area after irradiation. The hit efficiency calculated excluding this region is almost 99\% at 300~V for the 100~\mum{} thin sensors. The SCMs produced at CiS are still operative even after a very inhomogeneous irradiation. For the same irradiation fluence of 14$\times$10$^{15}$~\neqcm{} the 200 and 300~\mum{} thick sensors show similar collected charge up to a bias voltage of 1000~V. A first characterization of four chip modules before irradiation has been presented showing results comparable with those of the SCMs.

\acknowledgments
This work has been partially performed in the framework of the CERN RD50 Collaboration. The authors thank A.~Dierlamm (KIT), S.~Seidel (NMU), V.~Cindro, and I.~Mandic (Jo\v{z}ef-Stefan-Institut) for the sensor irradiations. The beam test measurements and part of the irradiations have received funding from the European Commission under the FP7 Research Infrastructures project AIDA, grant agreement no.~262025.

\end{document}